\def   \ni {\noindent}

\def   \sssk {\vskip  3truept}
\def   \ssk {\vskip  5truept}

\def   \bsk {\vskip 15truept}

\def   \newline {\hfil\break}

\documentstyle[epsfig]{article}
\begin{document}

\hsize 5truein
\vsize 8truein
\font\abstract=cmr8
\font\keywords=cmr8
\font\caption=cmr8
\font\references=cmr8
\font\text=cmr10
\font\affiliation=cmssi10
\font\author=cmss10
\font\mc=cmss8
\font\title=cmssbx10 scaled\magstep2
\font\alcit=cmti7 scaled\magstephalf
\font\alcin=cmr6 
\font\ita=cmti8
\font\mma=cmr8
\def\ref{\par\noindent\hangindent 15pt}
\null
\def\ol{\Omega_{\Lambda}}
\def\ob{\Omega_{b}}
\def\obh{\Omega_{b}h^{2}}
\def\oc{\Omega_{CDM}}
\def\om{\Omega_{m}}

\def\gtrsim{\mathrel{\hbox{\rlap{\hbox{\lower4pt\hbox{$\sim$}}}\hbox{$>$}}}}
\def\lesssim{\mathrel{\hbox{\rlap{\hbox{\lower4pt\hbox{$\sim$}}}\hbox{$<$}}}}
\def\lsim   {\wisk{_<\atop^{\sim}}}
\def\gsim   {\wisk{_>\atop^{\sim}}}

\def\etal{{\em et al.~}}
\def\apj{{\em Ap.J.}}
\def\apjl{{\em Ap.J.L.}}
\def\mnras{{\em M.N.R.A.S.}}
\def\aa{{\em A \& A}}
\def\be{\begin{equation}}
\def\ee{\end{equation}}

\def\thefirstfig{
\makebox{
\medskip
\noindent
\parbox[l]{1.5truein}{
\footnotesize
{\bf Figure 1. The Ambush}\\
These 6 panels show the details of the parameter space ambush
of $\ol$ and $\om$.
CMB constraints are in the top left panel.
``Conservative'' versions of constraints from double-lobed radio sources, 
Type Ia supernovae (``SN'') and galaxy cluster mass-to-light ratios (``M/L'')
are shown on the right along with their joint likelihoods with the CMB.
The three shades of grey are approximate 1, 2 and 3-$\sigma$ confidence regions.
More optimistic versions of the SN and M/L constraints are on the left.
The conservative versions on the right were used to make Figure 2.
The CMB excludes the lower left region of the $\om - \ol$ plane while each of the 
other constraints excludes the lower right. $\Lambda$-CDM models in the upper left
are the only models consistent with all the data sets.
}
\hglue0.1truein
\parbox[r]{3.3truein}{\epsfxsize=3.3truein\epsfbox{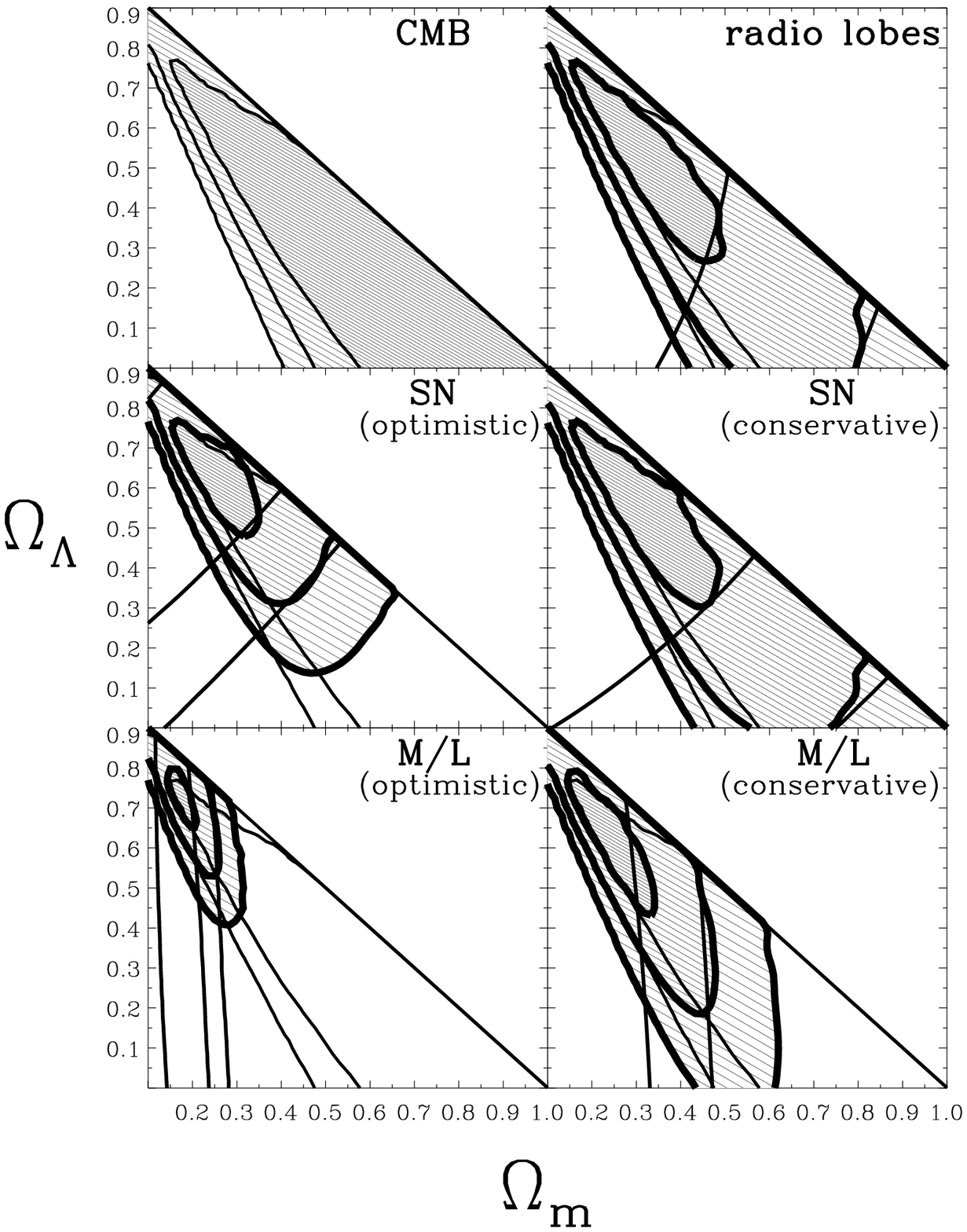}}
\smallskip
}
}
\def\thesecondfig{
\makebox{
\medskip
\noindent
\parbox[l]{1.4truein}{
\footnotesize
{\bf Figure 2. We've got you surrounded}\\
The combined constraints from CMB and non-CMB observations.
The combination of the three conservative versions from the right side of
Figure 1 yield the white contours ($\sim 68\%$ and $95\%$ confidence levels). The combination of the
elongated CMB triangle (from Figure 1, top left) with the white contours yields: 
$\ol = 0.62 \pm 0.16$ and $\om = 0.24 \pm 0.10$.
The black region in the upper right was not tested with CMB data.
}
\hglue0.1truein
\parbox[r]{3.3truein}{\epsfxsize=3.3truein\epsfbox{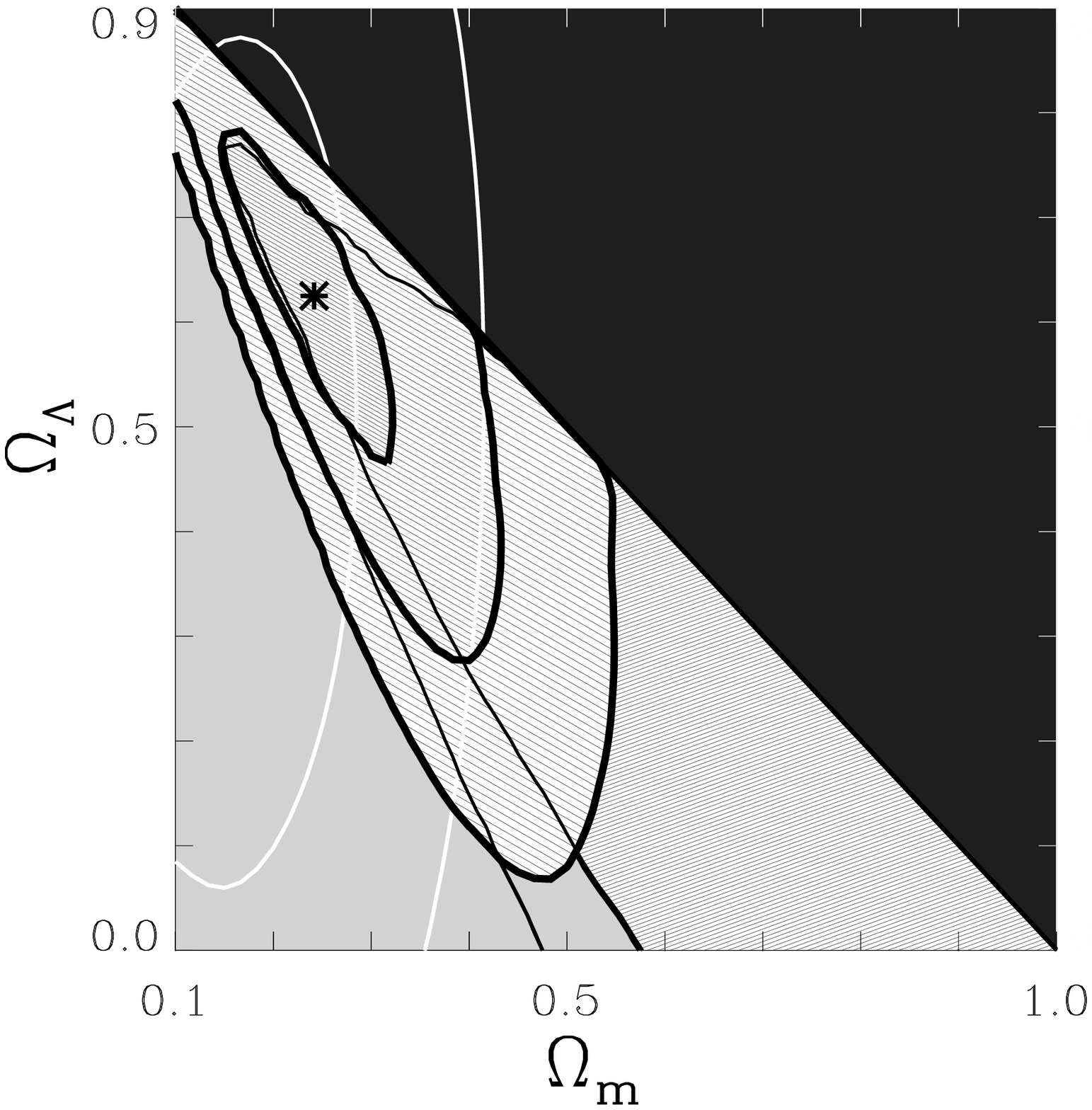}}
\smallskip
}
}
\def\thethirdfig{
\makebox{
\medskip
\noindent
\parbox[l]{1.4truein}{
\footnotesize
{\bf Figure 3. A line up of cosmological culprits}\\
$\ol$ is the big shot controling the Universe.
He's going to make it blow up.
$\oc$ would like to make the Universe collapse but
can't compete with $\ol$. $\ob$ just follows 
$\oc$ around.
Like all dangerous criminals, one can never be sure of 
$\ol$ until he is behind bars.
The CMB police is being beefed up.
Hundreds of heroic CMB observers are now planning his capture.
}
\hglue0.1truein
\parbox[r]{3.3truein}{\epsfxsize=3.3truein\epsfbox{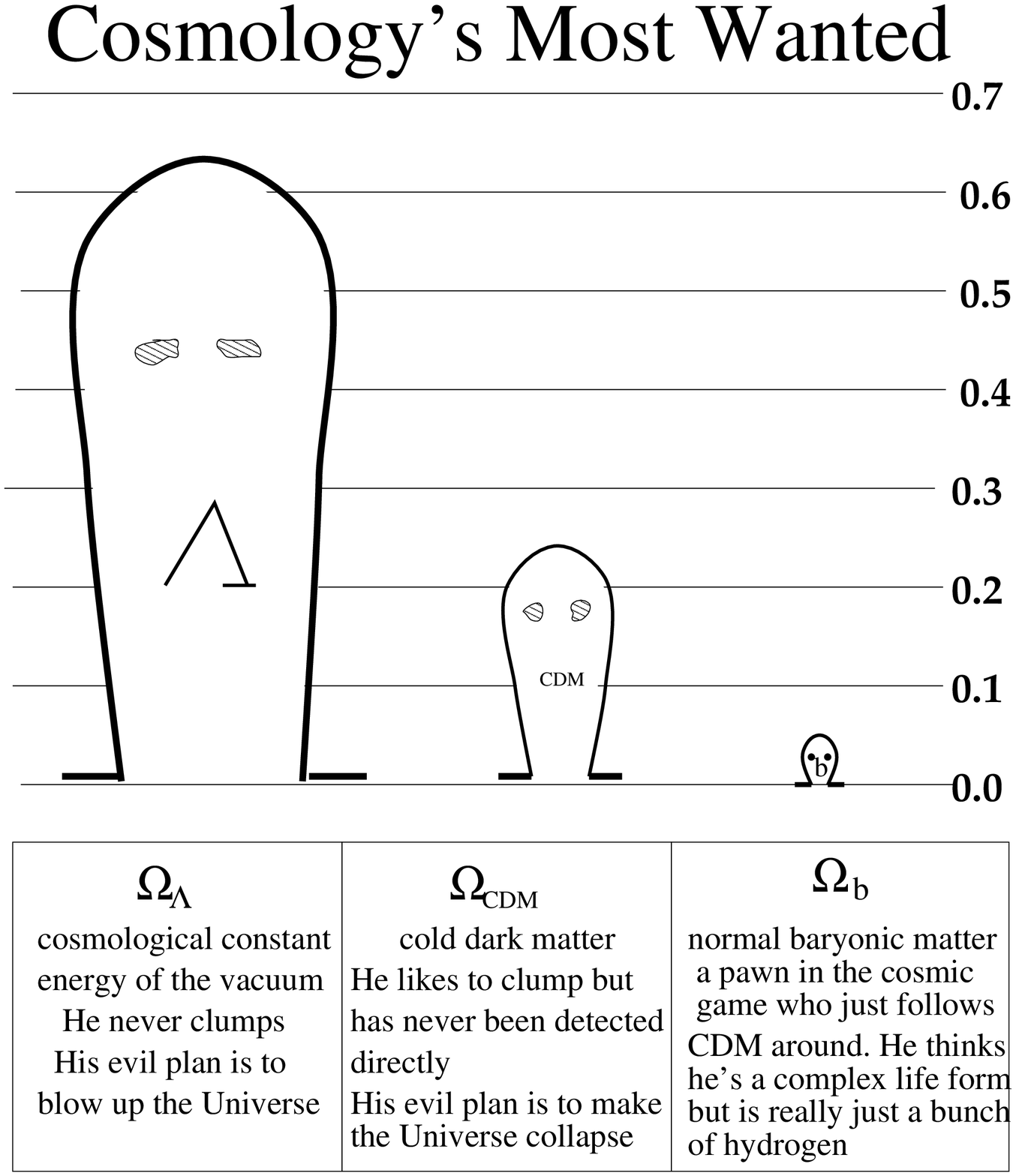}}
\smallskip
}
}


\title{\ni \begin{center} \boldmath $\displaystyle \ol, \: \om$ \unboldmath!\\
\sssk
 We know you're in there.\\
We've got you surrounded.\\
Come out with your hands up.
\end{center}
}  

\bsk \bsk
\author{\ni Charles H. Lineweaver}                                                       
\bsk
\affiliation{University of New South Wales, Sydney, Australia } 

\bsk
\baselineskip = 12pt

\abstract{ABSTRACT \ni
Recent cosmic microwave background (CMB) measurements  
combined with recent supernovae limits and other 
observational data have ambushed $\ol$ and $\om$. These cosmological culprits 
are trapped in a small pocket of parameter space: 
$\ol = 0.62 \pm 0.16$ and  $\om = 0.24 \pm 0.10$.
If this is correct then an unknown 2/3 of the Universe has been identified.

}                                                    
\bsk
\baselineskip = 12pt
\keywords{\ni KEYWORDS: cosmology; observational; cosmic microwave background radiation; cosmological 
constant
}               

\bsk
\baselineskip = 12pt



\text{\ni 1. WHY $\ol$ CANNOT BE ZERO
\ssk
\ni     
The combination of CMB and supernovae constraints provides the strongest evidence we have
that $\ol > 0$. If any $\ol = 0$ model can squeak by the new supernovae
constraints it is the very low $\om$ models.
However these models are the ones most strongly excluded by the CMB data.
The $\Lambda$-CDM region of the $\om - \ol$ plane does fit the CMB, supernovae and other 
data sets and should be viewed as the new standard model of cosmology.
Standard CDM with $\om = 1$ and $\ol = 0$ is a simpler model, but circular
planetary orbits are also simpler than ellipses.

Recent CMB anisotropy measurements favor the elongated triangle in the top left panel
of Figure 1. See Lineweaver (1998) for details. This plot shows that if $\ol = 0$ then $\om \sim 0.3$ is 
more than $\sim 4 \sigma$ from the best-fit and $\om \sim 0.1$ is more than $\sim 7 \sigma$
away. The N-$\sigma$ levels in this diagram are very rough but the message is clear:
if $\ol = 0$, then low $\om$ models are strongly excluded by the CMB data.
No other data set can exclude this region with such high confidence.
The CMB exclusion of the only $\ol = 0$ model permitted by generous supernovae constraints
is a fundamental result with far reaching implications about the nature of the
vacuum of quantum field theory.
Further, the combination of CMB constraints with supernovae and other cosmological
observations yields the most accurate determination of the cosmological constant
and thus the best evidence for the existence of more than 2/3 of 
the energy of the Universe.

\clearpage
\thefirstfig
\bsk

What is it about the current CMB data that excludes low $\om$ ($\ol= 0$) models?
Robust features of the power spectrum of the current CMB data set are: 
it's flat ($2 \lesssim \ell \lesssim 20$), it goes up ($30 \lesssim \ell \lesssim 200$)
and then it comes down ($300 < \ell < 700$). 
The position of the
peak is $\ell_{peak} = 260^{+30}_{-20}$ (this statement is slightly model
dependent and is strictly true only in $\ol = 0$ models, see Lineweaver \& Barbosa
1998).
Low $\om$ models peak at much larger $\ell$ values (smaller angular scales)
than the current data. For example if $\om \sim 0.2$ then the peak of the power spectrum
would have to be at  $\ell \sim 700$.
Saskatoon, CAT and OVRO observations in the region  $300 < \ell < 700$
all indicate that the spectrum is coming down in this region, not going up.
Also, the rise to the first peak in the region $30 \lesssim \ell \lesssim 200$
is probably the most robust feature of the data.
The observed rise is much larger over this interval than would be the case
in low $\om$ models.
So the position of the peak and the rise to the peak are robust
features of the current data and  both  are incompatible
with low $\om$ ($\ol = 0$) models.

}
\ssk
\text  {\ni 2. WE'VE GOT YOU SURROUNDED, ROBUSTLY
\ssk
\ni     

As more cosmological data comes in, the CMB and non-CMB constraints
form an ever-tightening network of interlocking constraints, thus building
the most complete and unified picture of the Universe on the largest 
scales.
Figure 1 shows some of the pieces of this ever-tightening network.
Recent CMB constraints are in the upper left panel.
In each of the other 5 panels, the joint likelihood of the CMB with another
observational constraint is shown.

The upper right panel shows constraints from observations at redshifts between 0 and 2 
of double-lobed radio sources which were
used as standard rulers (Daly \etal 1998).
Two competing supernovae groups are following the apparent magnitudes of
Type Ia supernovae at redshifts between 0 and 1 and are using these approximately standard
candles to put constraints in the $\om - \ol$ plane (Riess \etal 1998, Perlmutter \etal 1998).
Conservative and optimistic versions of these constraints are shown in the middle panels of
Figure 1.
Studies of the mass-to-light ratio in galaxy clusters at redshifts between $0$ and $0.6$ 
favor low values of $\om \sim 0.2$ and have only a 
slight $\ol$-dependence (Carlberg \etal 1997, 1998).
Conservative and optimistic versions of these ``M/L'' constraints are shown in
the bottom two panels of Figure 1.
Studies of cluster evolution yield slightly larger values: $\om \sim 0.3 \pm 0.1$ with only 
slight $\ol$-dependence (Bahcall \etal 1997).  
More details about the contours shown in Figure 1 can be found in
Table 1 of Lineweaver (1998).

Figure 2 shows the main result:
$\ol = 0.62 \pm 0.16$ and  $\om = 0.24 \pm 0.10$.
These numbers come from the joint likelihood of the CMB constraints (top left panel of Figure 1)
with the 3 non-CMB constraints in the 3 panels of the right side of Figure 1.
Although the main results quoted here are the best current limits in 
the $\om - \ol$ plane, I believe they are also robust because of a series of
conservative choices made in the analysis.
There are a variety of ways in which the limits can be selected and combined.
My strategy is to be reasonably conservative by trying not to over-constrain parameter space.
Practically this means using contours large enough to include possible systematic errors. 
For example, the two independent supernovae groups are in the process of taking data and 
refining their analysis and calibration techniques.
I use the ``SN conservative'' contours (Figure 1, middle right) which are large enough to 
subsume the worst case systematic errors
that either group has calculated (see Figures 6 and 7 of Riess \etal 1998 and Figure 5
of Perlmutter \etal 1998).
 
I apply the same strategy with the Carlberg \etal (1997, 1998) cluster mass-to-light
ratios. They report 30\% errors in their $\om$ result but also
cite a worst case 73\% error if all the systematic errors conspire and 
add linearly. I use the 73\% error in the ``M/L conservative'' contours (lower right panel of Figure 1).
For comparison the 30\% error is shown in the lower left panel of Figure 1
labeled ``M/L optimistic''.
An additional way in which the non-CMB contours in Figure 1 are conservative is that
I have taken published $\Delta \chi^{2} = 2.3$ contours
and treated them as $\Delta \chi^{2} = 1$ contours.
 
The results I report are also robust in the sense that
when combined individually (as in Figure 1) or combined together (as in Figure 2)
the result is much the same.
Systematic errors may compromise one or the other of the 
observations but are less likely to bias all of the 
observations in the same way.

\clearpage
\thesecondfig

We have tracked the cosmological culprits $\om$ and $\ol$ down into a small
region of parameter space.
CMB is forcing them to the right while SN are forcing them into the upper left.
They're caught in a cross-fire.
Cluster mass-to-light ratios (and cluster evolution) has got them pinned down 
in a vertical strip at  $\om \sim 0.2$ (or $0.3$).
They have no where else to go.
The CMB excludes low $\om$ open-CDM models while supernovae and other observations
exclude standard-CDM models.
$\Lambda$-CDM seems to be faring much better than standard-CDM at running
the gauntlet of cosmology tests.
This region should be the new standard model of cosmology.
{\bf Whatever the true model of the Universe is, it has to look alot like
$\Lambda$-CDM}.

}
\bsk
\text{\ni 3. HAVE WE GOT THE WRONG GUY?
\ssk
\ni     

A vocal minority of $\Lambda$-phobic cosmologists and particle
theorists believe that any $\Lambda > 0$ result has got to be wrong.
Their reasoning goes something like this. Theory predicts that $\ol \gtrsim 10^{52}$. Since it is 
obviously not this value, $\ol$ must be zero based on some principle we don't understand yet.
See Cohn (1998, Section II) for a more judicious discussion.

But how could the result reported here be wrong?
What are the weakest points in the analysis?
The CMB result is only as good as the models assumed.
Maybe the parameter space is too narrow or, more fundamentally,  maybe the models are just wrong.
We have looked at a popular but limited region of inflation-based gaussian CDM models with adiabatic 
initial conditions.  
If the goodness-of-fit of these models had been bad, we would have had good reason to be
dissatisfied. But the best-fit is also a good fit; almost too good:
$\chi^{2}= 22.1$ with 28 degrees of freedom. 
The probability of obtaining a lower value is only $22\%$.
Additionally when optical depth and tensor modes are added to the parameter
space the CMB exclusion of the lower left in Figure 2 does not change
(Tegmark 1998).

But maybe we've got the wrong guy.
The cosmological constant has $w=-1$ in the equation of state $p = w\rho$.
There are models of generalized dark matter or ``quintessence'' 
with values of $w \ne -1$ and models with $w$ as a function of time (Hu \etal 1998).
Recent results from supernovae have only a mild preference for
$w \sim -1$  models (Garnavich \etal 1998, Perlmutter \etal 1998),
so these more general models are still plausible.
See Cohn (1998, section VI) for discussion of this issue.

Not all data sets are happy with $\Lambda$-CDM.
Kochanek and collaborators find lensing limits which place limits
$\ol \lesssim 0.6$ in flat models (Kochanek 1996, Falco \etal 1998)
but other lensing limits agree with results reported here
(Fort \etal 1997, Chiba \& Yoshii 1997).
Peacock (1998) sees the shape of the APM power spectrum as in at least 
marginal conflict with $\Lambda$-CDM models but
the results reported in Webster \etal (1998) 
from a combination of CMB and IRAS $P(k)$ agree well with Figure 2.
Even velocity field data are happier with $(\om, \ol)= (0.3,0.7)$ than
with $(0.3, 0.0)$  (Dekel 1998, Zehavi 1998).

}
\ssk
\thethirdfig
\clearpage
\text{\ni 4. RESISTANCE IS FUTILE
\ssk
\ni     

As the quality and quantity of cosmological data improve 
questions about cosmological parameters will get increasingly precise answers
from an increasingly tight network of constraints.
Better CMB detectors are being built, long duration balloons will fly, 
sensitive new high resolution interferometers will soon be on line and 
we all have high expectations for the
two CMB satellites  MAP and Planck.
See Lasenby \etal (1998) for a recent review.

}
\ssk

\bsk
\baselineskip = 12pt
{\abstract \ni ACKNOWLEDGMENTS
I thank the organizers for organizing a meeting lively enough to compete
with the World Cup and some beautiful sunny beaches.
I am supported by at UNSW by a Vice Chancellor's Research Fellowship.
}

\bsk
\baselineskip = 12pt


{\references \ni REFERENCES
\ssk

\ref Bahcall, N. Fan, X. \& Cen, R. 1997, \apj 485, L53
\ref Carlberg, R.G. \etal 1997, \apj 478, 462
\ref Carlberg, R.G. \etal 1998 in Proc. 33rd Moriond Astrophysics Meeting,
Fundamental Parameters in Cosmology, ed. J.Tran Thanh Van \& Y. Giraud-Herauld (Gif-sur-Yvette:
Editions Frontieres), in (astro-ph/9804312).
\ref Chiba, M. \& Yoshii, Y. 1997, \apj 490, L73
\ref Cohn, J. 1998, astro-ph/9807128
\ref Daly, R.A. Guerra, E.J. \& Wan, L. 1998, in Proc. 33rd Moriond Astrophysics
Meeting: Fundamental Parameters in Cosmology, ed. J. Tran Thanh Van \& Giraud-Heraud 
(Gif-sur-Yvette, Editions Frontieres, in press, astro-ph/9804312)
\ref Dekel, A. 1998, private communication
\ref Falco, E.E., Kochanek, C.S. \& Munoz, J.A. 1998, \apj 494, 47 
\ref Fort, B. \etal 1997, A\&A 321,353
\ref Garnavitch, P.M. \etal 1998, \apj 509, in press
\ref Hu, W. \etal, 1998, astro-ph/9806362
\ref Kochanek, C.S. 1996, \apj 466, 638
\ref Lasenby, A.N. \etal 1998, astro-ph/9810196
\ref Lineweaver, C. H. \& Barbosa, D. 1998, \apj 496, 624
\ref Lineweaver, C. H. 1998, \apj 505, L69
\ref Peacock, J.A. 1998, Phil. Trans. R. Soc. London A, submitted astro-ph/9805208
\ref Perlmutter, S. \etal 1998, in preparation  
\ref Riess, A.G. \etal 1998, A.J. in press, astro-ph/9805201
\ref Tegmark, M. 1998, \apj submitted, astro-ph/9809201
\ref Webster, M. \etal 1998, \apj submitted, astro-ph/9802109
\ref Zehavi, I. 1998, in Proc. of Evolution of Large Scale Structure, Garching,
astro-ph/9810246

}                      

\end{document}